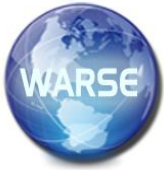

# Offshore Software Maintenance Outsourcing: Predicting Client's Proposal using Supervised Learning

**Atif Ikram[1,2], Masita Abdul Jalil[1], Amir Bin Ngah[1], Ahmad Salman Khan[3], Tahir Iqbal[4]**

[1]School of Informatics and Applied Mathematics, Universiti Malaysia Terengganu, Malaysia
[2]Department of Computer Science & Information Technology, The University of Lahore, Pakistan
[3]Department of Software Engineering, The University of Lahore, Pakistan
[4]Department of Computer Science, Bahria University, Lahore Campus, Pakistan
aikram4u@gmail.com, p4086@pps.umt.edu.my, atif.ikram@cs.uol.edu.pk

## ABSTRACT

In software engineering, software maintenance is the process of correction, updating, and improvement of software products after handed over to the customer. Through offshore software maintenance outsourcing (OSMO) clients can get advantages like reduce cost, save time, and improve quality. In most cases, the OSMO vendor generates considerable revenue. However, the selection of an appropriate proposal among multiple clients is one of the critical problems for OSMO vendors. The purpose of this paper is to suggest an effective machine learning technique that can be used by OSMO vendors to assess or predict the OSMO client's proposal. The dataset is generated through a survey of OSMO vendors working in a developing country. The results showed that supervised learning-based classifiers like Naïve Bayesian, SMO, Logistics apprehended 69.75 %, 81.81 %, and 87.27 % testing accuracy respectively. This study concludes that supervised learning is the most suitable technique to predict the OSMO client's proposal.

## 1. INTRODUCTION

Software maintenance is the longest lifetime and budget consuming phase of software systems as it consumes more than 70 % of the total allocated budget of the software development lifecycle. The trend of maintenance outsourcing is increasing among software companies to achieve quality while saving time and money. In offshore software maintenance outsourcing (OSMO), the OSMO vendor provides (usually from a developing country) the required software maintenance services to OSMO clients (usually from a developed country) [1]. Although the client's proposal is giving business even then the OSMO vendor should carefully accept the OSMO client's proposal, as an aphorism "All that glitters is not gold". The OSMO vendor can select an appropriate or more suitable proposal among multiple options, with the help of some prediction or assessment-related techniques. It is of great importance to assess, predict, or estimate the client's proposal before its acceptance. Therefore, there exist several studies that use estimation or machine learning techniques on this subject [2], [3], [4], [5].

In machine learning, supervised learning is a classification learning advent that is used to evaluate training or label data in an attempt to model hidden and unseen data for future and inevitable classification [6]. There exist several studies which endorse the use of machine learning for assessment or prediction purposes in different domains. Malhotra and Chug (2016) [2] highlighted the importance of prediction of software maintainability. The study discussed that the use of machine learning algorithms for the prediction of software maintenance has been significantly increased since 2005. This study does not cover an offshore context. The study [3] has encouraged vendors to use machine learning techniques like Neural Network to assess the client's proposal at the project selection stage. The study concludes that the use of machine learning techniques will help the vendor's project managers to select the most suitable project among many others. The studies [8] and [9] have identified several factors that can be used as causal agents to predict the OSMO client's proposal. The studies have a good contribution towards offshore software outsourcing business but have limitations that their focus is on software development outsourcing instead of maintenance outsourcing.

The current study has used attributes (variable) as casual agents in the prediction process of OSMO client's proposal as used by studies [10] and [11] in the prediction of software project success or failure. The studies [12], [13],[14], [15],[16],[17],[18],[24],[25] and [26] have identified 17 attributes. Table 1 shows these 16 attributes or casual agents.

The current study will use machine learning techniques or supervised learning classifiers along with casual agents to evaluate the OSMO client's proposal. This prediction will help the OSMO vendor in the assessment of the OSMO client's proposal and will guide him to accept or reject this business offer. Therefore, the current study has three research questions to answer:

RQ1: What are the key variables or attributes that can impact the OSMO client's proposal?
RQ2: Which is the most appropriate technique to predict the OSMO client's proposal?
RQ3: Which is the most accurate SL classifier to predict the OSMO client's proposal?

. **Table 1:** Attributes Identified

| Sr. | Attribute | Sr. | Attribute |
|---|---|---|---|
| 1 | Size of the supplier organization | 10 | Structure of code (good, avg, poor ) |
| 2 | Required team size | 11 | Common time zone |
| 3 | Domain of the project | 12 | Client's market reputation |
| 4 | Size of software maintenance project | 13 | Handover experience of the client (software and knowledge) |
| 5 | Use of international standard in the software development phase | 14 | Operating language (client's language) (similar or other) |
| 6 | Software code complexity | 15 | Nature of SLA (fair, unclear, biased) |
| 7 | Use of international standard in software documentation | 16 | Methodology adopted |
| 8 | Quality of related document | 17 | System Age (newly developed, Legacy system) |
| 9 | Required type of maintenance | Total = 17 | |

## 2. PROPOSED SCHEME

The research method used in this study is mixed [19]. The mixed-method combines qualitative (interview, case study) and quantitative (use of statistical or mathematical techniques) research methods. The study [20] advocates that researchers can gain considerable benefits by using the mixed method in the field of software projects. To reach the final solution, the study is conducted in the following steps shown in Figure 1. The detail of every step is discussed further.

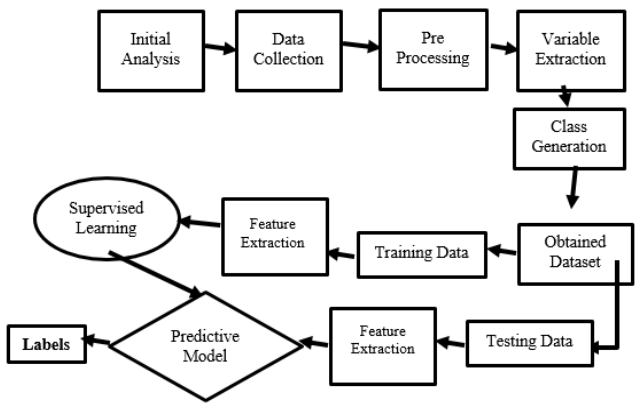

**Figure 1**: The proposed model for the current study

### 2.1 Initial Analysis

The OSMO vendor required a more reliable technique for the prediction of OSMO's client's business proposals. In this phase of the study, the research model is proposed and different methodologies and attributes are reviewed in detail. This preliminary analysis indicated some fundamental independent variables (Table 1) which lead towards providing an intelligent and optimized supervised learning-based solution for better predictions of OSMO proposals.

### 2.2 Data Collection

Data collection is the process of gathering and measuring information on variables of interest, in an established systematic relationship that enables to answer research questions, test hypotheses, and evaluate outcomes. The multivariate analysis deals with the statistical analysis of data collected on more than one dependent variable. Multivariate techniques are popular because they help organizations to turn data into knowledge and thereby improve their decision making. Use letters for table footnotes (see Table I).

In this phase of the study, a questionnaire is prepared for analysis to frame collected data on dependent and independent variables of the study. The questioner was based on structured (nominal scale) and unstructured questions. The questionnaire was sent to 50 software and IT companies of Pakistan. In reply, the current research received data of 173 software maintenance outsourcing projects.

### 2.3 Pre-processing

To improve the accuracy of the model, the training set must be complete, continuous, and noiseless. Therefore, it is necessary to pre-process the original data. Pre-processing includes data cleaning, data integration, data conversion, and data simplification. Pre-processing is a process of inspecting, cleansing, transforming, and modeling data. The response recorded from different companies had certain major issues like missing values, redundancy, etc. After the pre-processing phase, 165 responses were considered for the further phases of the study.

### 2.4 Variable Extraction

The detail of the variables extracted from the collected data with description is illustrated in Table 2.

**Table 2**: Extracted Variables and Nominal Scale Values

| S.No | Variable Name | Nominal Scale Vales |
|------|---------------|---------------------|
| 1 | Supplier_Size | Small, Medium, Large |
| 2 | Required_Team | Small, Medium, Large |
| 3 | Domain | Same, Partial_Same, New_Domain |
| 4 | Project_Size | Small, Medium, Large |
| 5 | Development_Standard | Fully, Partial, No |
| 6 | Code_Complexity | Easy, Challenging |
| 7 | Document_Standard | Fully, Partial, No |
| 8 | Document_Quality | Good, Average, Poor |
| 9 | Maintenance | Single, Multiple |
| 10 | Code_Structure | Good, Average, Poor |
| 11 | Common_Time | Fully, Partial, No |
| 12 | Client_Repute | Good, Average, Poor |
| 13 | Client_Experience | Exp, Semi-Exp, New_Cus |
| 14 | Operating_Language | English_as_Comm, Different Lang |
| 15 | SLA_Nature | Fair, Biased, Unclear |
| 16 | Method_Adopted | Waterfall, Iterative, Agile, |
| 17 | System_Age | Newly_Developed, Legacy |

### 2.5 Class Generation

The instances of the dataset can be classified into three label classes as **Accept, Risk** and **Reject**.

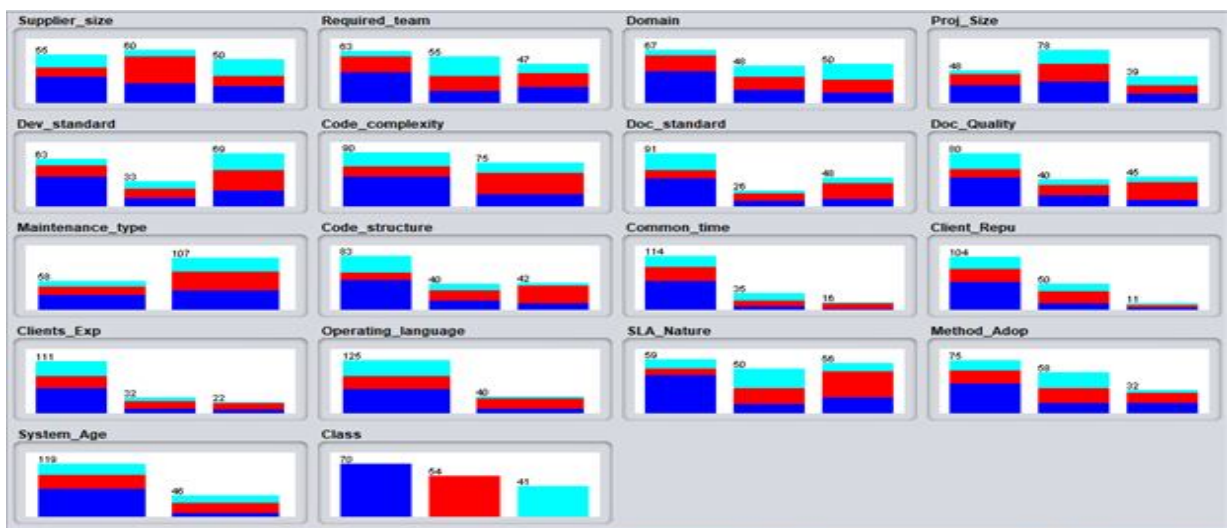

**Figure 1**: Distribution of Attributes for Label Classes

## 2.6 Obtained Dataset

The dataset obtained consists of 165 instances with 17 nominal attributes where each instance belongs to one label class. For this study, the dataset of 165 instances is divided into 66 % of the training set (110 instances) and 34 % testing set (55 instances). The detailed description of attributes and classes is already discussed in Table 1 and Table 2.

Few dataset instances are illustrated here as a sample where **'Accept', 'Risk'** and **'Reject'** are label classes:
**1**.medium,small,partial_same,medium,partially,easy,partially,avg,multiple,avg,partially,avg,sami_exp,English_as_Common,Fair,Waterfall,Legacy,
**Accept**
**2**.medium,small,Same,large,no,challenging,fully,good,single,avg,fully,avg,new_customer,English_as_Common,Unclear,Iterative,Legacy,**Risk**
**3**.medium,large,new_domain,large,partially,easy,fully,good,multiple,good,fully,good,new_customer,English_as_Common,Fair,Agile,Newely_dev, **Reject**

The distribution of attributes with respect to label classes is visualized in Figure 1.

**Table 3**: Information Gain Value of the Attributes

| Rank | Variable Name | Information Gain |
|------|---------------|------------------|
| 1 | Expected_SLA_Nature | 0.1921 |
| 2 | Code_Structure | 0.1559 |
| 3 | Common_Time | 0.1225 |
| 4 | Document_Standard | 0.1147 |
| 5 | Document_Quality | 0.1075 |
| 6 | Supplier_Size | 0.0891 |
| 7 | Domain | 0.0882 |
| 8 | Code_Complexity | 0.0879 |
| 9 | Required_Team | 0.0856 |
| 10 | Method_Adopted | 0.0832 |
| 11 | Operating_Language | 0.0797 |
| 12 | System_Age | 0.0774 |
| 13 | Development_Standard | 0.0723 |
| 14 | Client_Experince | 0.0649 |
| 15 | Clients_Repute | 0.0634 |
| 16 | Project_Size | 0.0344 |
| 17 | Maintainance_Type | 0.0138 |

### 2.6.1 Training Results

WEKA™ (Waikato Environment for Knowledge Analysis) version 3.9 which is considered one of the most efficient tools for machine learning and data mining, has been used to find the overall efficiency of the proposed technique and the dataset.

Different categories of algorithms are available for supervised learning, like Bayes, Function, Lazy, Meta, Rules, and Tree, etc. In this study, the efficiency of the proposed technique and the dataset is demonstrated by using classifiers like Bayes, Function, Rules, Tree, etc.

The Bayesian network consists of a structural model and a set of conditional probabilities. Bayesian-based algorithms are often used for classification problems in which learning is done by constructing a classifier from a set of training instances with labeled classes [21].

Bayes algorithms are simple supervised probabilistic classifier algorithms [23] used for binary or multiclass classification based on Bayes Theorem also these algorithms are highly scalable. Naïve Bayes algorithm is a simple supervised probabilistic classifier algorithm used for binary or multiclass classification based on Bayes Theorem. This algorithm is highly scalable. The obtained training results generated by different supervised learning algorithms are shown in Table 4.

**Table 4**: Obtained Training Results Generated by Different Supervised Learning Algorithms

| Supervised Learning Algorithms | | Training Results Parameters | | | | | | | |
|---|---|---|---|---|---|---|---|---|---|
| | | No of Inst | Correctly Classified Instances | Incorrectly Classified Instances | Kappa Statistic | Mean Absolute Error | Root Mean Squared Error | Relative Absolute Error | Root Relative Squared Error |
| Bayes | Bayes Net | 110 | 80 (72.72 %) | 30 (27.27 %) | 0.571 | 0.192 | 0.381 | 45.53 % | 82.40 % |
| | Naïve Bayes | 110 | 80 (72.72 %) | 30 (27.27 %) | 0.573 | 0.197 | 0.380 | 45.82 % | 82.14 % |
| Function | Logistics | 110 | 106 (96.36 %) | 4 (3.63 %) | 0.943 | 0.070 | 0.177 | 16.49 % | 38.36 % |
| | Multi-layer Perceptron | 110 | 105 (95.45 %) | 5 (4.545%) | 0.929 | 0.046 | 0.147 | 10.88 % | 31.73 % |
| | SMO | 110 | 99 (90.00 %) | 11 (10.00 %) | 0.843 | 0.256 | 0.329 | 59.66 % | 71.04 % |
| Lazy | IBK | 110 | 108 (98.18 %) | 2 (1.818 %) | 0.971 | 0.025 | 0.089 | 6.00 % | 19.30 % |
| | K-Star | 110 | 108 (98.18 %) | 2 (1.818 %) | 0.971 | 0.021 | 0.091 | 5.087 % | 19.77 % |
| | Lazy-LWL | 110 | 85.45 (65 %) | 16 (14.54 %) | 0.775 | 0.270 | 0.322 | 62.94 % | 69.63 % |
| Meta | Bagging | 110 | 91 (82.72 %) | 19 (17.27 %) | 0.724 | 0.255 | 0.314 | 59.43 % | 67.81 % |
| | Classification via Regression | 110 | 95 (86.36%) | 15 (13.63 %) | 0.785 | 0.190 | 0.275 | 44.29 % | 59.37 % |
| | Iterative Classifier Optimizer | 110 | 91 (82.72 %) | 19 (17.27 %) | 0.725 | 0.225 | 0.313 | 52.39 % | 67.54 % |
| | Logit Boost | 110 | 93 (84.54 %) | 17 (15.45 %) | 0.754 | 0.215 | 0.303 | 49.99 % | 65.37 % |
| | Multi Class Classifier | 110 | 99 (90.00 %) | 11 (10.00 %) | 0.843 | 0.124 | 0.241 | 28.96 % | 52.02 % |
| | Randomizable Filtered Classifier | 110 | 108 (98.18 %) | 2 (1.818 %) | 0.97 | 0.025 | 0.089 | 6.00 % | 19.30 % |
| Rules | Part | 110 | 92 (83.63 %) | 18 (16.36 %) | 0.740 | 0.176 | 0.297 | 41.07 % | 64.11 % |
| Tree | Hoeffding Tree | 110 | 80 (72.72 %) | 30 (27.27 %) | 0.573 | 0.197 | 0.380 | 45.82 % | 82.14 % |
| | LMT | 110 | 91 (82.72 %) | 19 (17.27 %) | 0.725 | 0.225 | 0.313 | 52.39 % | 67.54 % |
| | Random Forest | 110 | 108 (98.18 %) | 2 (1.818 %) | 0.971 | 0.097 | 0.149 | 22.75 % | 32.22 % |
| | Random Tree | 110 | 108 (98.18 %) | 2 (1.818 %) | 0.971 | 0.015 | 0.088 | 3.66 % | 19.15 % |

**Table 5**: Obtained Testing Results Generated by Different Supervised Learning Algorithms

| Supervised Learning Algorithms | | Testing Results Parameters | | | | | | | |
|---|---|---|---|---|---|---|---|---|---|
| | | No of Inst | Correctly Classified Instances | Incorrectly Classified Instances | Kappa Statistic | Mean Absolute Error | Root Mean Squared Error | Relative Absolute Error | Root Relative Squared Error |
| **Bayes** | Bayes Net | 55 | 35 (65.45%) | 20 (34.54%) | 0.452 | 0.262 | 0.460 | 59.69 % | 97.23 % |
| | Naïve Bayes | 55 | 38 (69.75 %) | 17 (30.90 %) | 0.512 | 0.232 | 0.430 | 53.32 % | 91.95 % |
| **Function** | Logistics | 55 | 48 (87.27 %) | 7 (13.73 %) | 0.166 | 0.362 | 0.594 | 18.53 | 24.17 % |
| | Multi-layer Perceptron | 55 | 45 (81.81 %) | 10 (19.18%) | 0.297 | 0.311 | 0.517 | 70.83 % | 19.13% |
| | SMO | 55 | 45 (81.81 %) | 10 (19.18%) | 0.262 | 0.379 | 0.479 | 86.37% | 21.28% |
| **Lazy** | IBK | 55 | 35 (65.45 %) | 20 (34.54%) | 0.441 | 0.2576 | 0.464 | 58.59  % | 98.03 % |
| | K-Star | 55 | 37 (67.27 %) | 18 (32.72 %) | 0.497 | 0.258 | 0.445 | 58.76 % | 93.99 % |
| | Lazy-LWL | 55 | 30 (54.54 %) | 25 (45.45 %) | 0.305 | 0.37 | 0.447 | 84.15 % | 94.53 % |
| **Meta** | Bagging | 55 | 34 (61.81 %) | 21 (38.18 %) | 0.400 | 0.351 | 0.423 | 79.86 % | 89.28 % |
| | Classification via Regression | 55 | 29 (52.72 %) | 26 (47.27 %) | 0.262 | 0.345 | 0.453 | 78.56 % | 95.64 % |
| | Iterative Classifie Optimizer | 55 | 31 (56.36 %) | 24 (43.63 %) | 0.318 | 0.334 | 0.457 | 76.11 % | 96.59 % |
| | Logit Boost | 55 | 30 (54.54 %) | 25 (45.45 %) | 0.288 | 0.340 | 0.471 | 77.35 % | 99.50 % |
| | Multi-Class Classifier | 55 | 29 (52.72 %) | 26 (47.27 %) | 0.264 | 0.318 | 0.508 | 72.40 % | 107.2 % |
| | Randomizable Filtered Classifier | 55 | 31 (56.36 %) | 24 (43.63 %) | 0.315 | 0.298 | 0.527 | 67.98 % | 111.4 % |
| **Rules** | Part | 55 | 30 (54.54 %) | 25 (45.45 %) | 0.309 | 0.342 | 0.487 | 77.89 % | 102.8 % |
| **Tree** | Hoeffding Tree | 55 | 36 (65.45 %) | 19 (34.54 %) | 0.452 | 0.266 | 0.460 | 60.51 % | 97.18 % |
| | LMT | 55 | 31 (56.36 %) | 24 (43.63 %) | 0.318 | 0.334 | 0.457 | 76.11 % | 96.59 % |
| | Random Forest | 55 | 35 (65.45 %) | 20 (34.54%) | 0.437 | 0.303 | 0.408 | 68.91 % | 86.16 % |
| | Random Tree | 55 | 32 (58.18 %) | 23 (41.81%) | 0.350 | 0.287 | 0.521 | 65.42 % | 110.0 % |

**2.6.2 Testing Results**

The obtained testing results generated by different supervised learning algorithms are shown in Table 5

## 5. CONCLUSION

The testing accuracy achieved by different supervised learning classifiers like Naïve Bayes, SMO, Logistics, and Random Forest reached up to 72.72 %, 90.0 %, 96.36 %, and 98.18 % respectively. Other supervised learning algorithms like LMT, Logit Boost, and Iterati ve Classifier Optimizer, etc, are also showing promising results. The testing accuracy which depicts the true accuracy achieved by the proposed supervised learning-based classifiers like Naïve Bayes, SMO, and Logistics achieved 69.75 %, 81.81 %, and 87.27 % test results.

We studied the performance of different classifiers over our dataset and found that two classifiers SMO and Logistics had relatively high performance. Of these two, we prefer the Logistics classifier for its simplicity in terms of the number of attributes required to make a good prediction. Using Information Gain (Table 3), we found that we were able to generally improve the performance of the classifiers over that dataset. We also found a list of attributes deemed to be the most important by the Information Gain and that produced the best performance for the Logistics classifier. Concerning the research questions of the study, we found that:

RQ1. The top five key variables or attributes that can impact the OSMO client's proposal are SLA Nature, Code Structure, Common Time, Document Standard, and Document Quality. The least Information Gain is obtained from the variable Maintenance Type. The vendor should focus his efforts on these areas of the OSMO client's proposal to select a good client.

RQ2. The Supervised Learning technique is the most appropriate while predicting the OSMO client's proposal.

RQ3. The most appropriate classifier for predicting the OSMO client's proposal with this collection of data was Logistics.

## 6. FUTURE WORK AND ACKNOWLEDGEMENT

The authors faced difficulties and have limitations while collecting data from software houses. For machine learning techniques, high volume of data is more appreciated. Therefore, we recommend researchers to implement some other techniques like Fuzzy logic, Multi-Criteria Decision-Making (MCDM) etc. on the current study.

We are thankful to Universiti Malaysia Terengganu, Malaysia for providing state of the art research facilities.

## REFERENCES

1. Ikram, A., Jalil, M. A., Ngah, A. B., & Khan, A. S. (2020). Towards Offshore Software Maintenance Outsourcing Process Model, IJCSNS, 20(4), 6.
2. Malhotra, R., & Chug, A. (2016). Software maintainability: Systematic literature review and current trends. International Journal of Software Engineering and Knowledge Engineering, 26(08), 1221-1253..
3. Costantino, Francesco, Giulio Di Gravio, and Fabio Nonino. "Project selection in project portfolio management: An artificial neural network model based on critical success factors." International Journal of Project Management 33.8 (2015): 1744-1754.
4. Zhou.L, Pan.S, Wang.J, Vasilakos, "Machine Learning on Big Data: Opportunities and Challenges", Neurocomputing, (2017), 1-29.
5. Alsolai, H., & Roper, M. (2020). A systematic literature review of machine learning techniques for software maintainability prediction. Information and Software Technology, 119, 106214.
6. Rizvi.S.S.R, Abbass.S, Khan.A, Asad.M "Optical Character Recognition System for Nastalique Urdu-Like Script Languages Using Supervised Learning", International Journal of Pattern Recognition and Artificial Intelligence, 33, (2019), 1-32
7. Costantino, Francesco, Giulio Di Gravio, and Fabio Nonino. "Project selection in project portfolio management: An artificial neural network model based on critical success factors." International Journal of Project Management 33.8 (2015): 1744-1754
8. Khan, Siffat Ullah, and Abdul Wahid Khan. "Critical challenges in managing offshore software development outsourcing contract from vendors' perspectives." IET software 11.1 (2017): 1-11
9. Khan, Siffat Ullah, Mahmood Niazi, and Rashid Ahmad. "Factors influencing clients in the selection of offshore software outsourcing vendors: An exploratory study using a systematic literature review." Journal of systems and software 84.4 (2011): 686-699
10. Cerpa, Narciso, et al. "Evaluating logistic regression models to estimate software project outcomes." Information and Software Technology 52.9 (2010): 934-944
11. Reyes, Francisco, et al. "The optimization of success probability for software projects using genetic algorithms." Journal of Systems and Software 84.5 (2011): 775-785.
12. Grolinger, Katarina, et al. "EEF-CAS: An Effort Estimation Framework with Customizable Attribute Selection." International Journal of Advancements in Computing Technology 5.13 (2013): 1-14.
13. Westner, Markus, and Susanne Strahringer. "Determinants of success in IS offshoring projects: Results from an empirical study of German companies." Information & management 47.5-6 (2010): 291-299.
14. Ulziit, B., Warraich, Z. A., Gencel, C., & Petersen, K. (2015). A conceptual framework of challenges and solutions for managing global software maintenance. Journal of Software: Evolution and Process, 27(10), 763-792.
15. Shukla, R., Shukla, M., Misra, A. K., Marwala, T., & Clarke, W. A. (2012, June). Dynamic software maintenance effort estimation modeling using neural network, rule engine and multi-regression approach. In International Conference on Computational Science and Its Applications (pp. 157-169). Springer, Berlin, Heidelberg.

16. Mihalache, M., & Mihalache, O. R. (2016). A decisional framework of offshoring: Integrating insights from 25 years of research to provide direction for future. Decision Sciences, 47(6), 1103-1149.

17. Khan, Siffat Ullah, Mahmood Niazi, and Rashid Ahmad. "Factors influencing clients in the selection of offshore software outsourcing vendors: An exploratory study using a systematic literature review." Journal of systems and software 84.4 (2011): 686-699.

18. Gopal, Anandasivam, and Balaji R. Koka. "The role of contracts on quality and returns to quality in offshore software development outsourcing." Decision sciences 41.3 (2010): 491-516.

19. Fetters, M. D., Curry, L. A., & Creswell, J. W. (2013). Achieving integration in mixed methods designs—principles and practices. Health services research, 48(6pt2), 2134-2156.

20. Ahmedshareef, Zana, Miltos Petridis, and Robert T. Hughes. "The affordances of mixed method in software project management research." International Journal of Multiple Research Approaches 8.2 (2014): 201-220

21. Jiang.L, Li.C, Wang.S, Zhang.L "Deep Feature Weighting for Naïve Bayes and its Application to Text Classification" Engineering Application of Artificial Intelligence, (2016), pp: 26-39

22. Bielza, C., & Larrañaga, P. (2014). Discrete Bayesian network classifiers: a survey. ACM Computing Surveys (CSUR), 47(1), 1-43.

23. Ahmed, I., Ali, R., Guan, D., Lee, Y. K., Lee, S., & Chung, T. (2015). Semi-supervised learning using frequent itemset and ensemble learning for SMS classification. Expert Systems with Applications, 42(3), 1065-1073.

24. Ikram, A., Jalil, M. A., Ngah, A. B., & Khan, A. S. (2020). Critical Factors In Selection Of Offshore Software Maintenance Outsourcing Vendor: A Systematic Literature Review, Journal of Theoretical and Applied Information Technology, ISSN 1992-8645 98(18).

25. Rizvi, S. S. R., Khan, M. A., Abbas, S., Asadullah, M., Anwer, N., & Fatima, A. (2020). Deep Extreme Learning Machine-Based Optical Character Recognition System for Nastalique Urdu-Like Script Languages. The Computer Journal.

26. Ikram, A., Riaz, H., & Khan, A. S. (2018). Eliciting Theory of Software Maintenance Outsourcing Process: A Systematic Literature Review. INTERNATIONAL JOURNAL OF COMPUTER SCIENCE AND NETWORK SECURITY, 18(4), 132-143.